\documentclass[prb,reprint,notitlepage,groupedaddress,floatfix,superscriptaddress]{revtex4-2}
\usepackage[english]{babel}
\usepackage[T1]{fontenc}
\usepackage[final]{graphicx}
\usepackage{natbib}
\usepackage{hyperref}
\usepackage{amssymb}
\usepackage{amsmath}
\usepackage{mathrsfs}
\usepackage{xcolor}
\usepackage{bm}
\usepackage{siunitx}

\newcommand{\AK}[1]{{\color{blue}#1}}

\newcommand{\addMisha}[1]{\textcolor{magenta}{#1}}

\usepackage[normalem]{ulem}

\makeatletter

\begin{document}
\title{Theory of biexciton-polaritons in transition metal dichalcogenide monolayers}

\author{Andrey Kudlis}
\affiliation{Science Institute, University of Iceland, Dunhagi 3, IS-107, Reykjavik, Iceland}
\affiliation{Abrikosov Center for Theoretical Physics, MIPT, Dolgoprudnyi, Moscow Region 141701, Russia}
\email{andrewkudlis@gmail.com}
\author{Ivan A. Aleksandrov}
\affiliation{Department of Physics, Saint Petersburg State University, Universitetskaya Naberezhnaya 7/9, Saint Petersburg 199034, Russia}
\affiliation{Ioffe Institute, 194021 Saint Petersburg, Russia}
\email{i.aleksandrov@spbu.ru}
\author{Mikhail M. Glazov}
\affiliation{Ioffe Institute, 194021 Saint Petersburg, Russia}
\author{Ivan A. Shelykh}
\affiliation{Science Institute, University of Iceland, Dunhagi 3, IS-107, Reykjavik, Iceland}

\date{\today}

\begin{abstract}

We theoretically investigate a nonlinear optical response of a planar microcavity with an embedded transition metal dicalcogenide monolayer when the energy of a biexcitonic transition is brought in resonance with the energy of a cavity mode. We demonstrate that the emission spectrum of this system strongly depends on an external pump. For small and moderate pumps, we reveal the presence of a doublet in the emission with the corresponding Rabi splitting scaling as a square root of the number of the excitations in the system. Further increase of the pump leads to the reshaping of the spectrum, which demonstrates, at weak damping, the pattern akin a Mollow triplet. An intermediate pumping regime shows a broad irregular spectrum reminiscent of a chaotic dynamics of the system.
\end{abstract} 

\maketitle

\section{Introduction}\label{sec:intro}
Transition metal dichalcogenides (TMDs) is a novel class of truly two-dimensional (2D) materials with fascinating optical response, dominated by bound electron-hole complexes, formed due to the strong Coulomb attraction between photocreated carriers. The simplest example of such a complex is an exciton, a bound state of an electron-hole pair, representing a solid state analogue of a hydrogen atom~\cite{Mak:PRL105(2010),Splendiani:NanoLett10(2010),Chernikov:PRL115(2015),Ugeda:NatMat:13(2014),He:PRL113(2014),Zhang:PRB89(2014),Singh:PRB93(2016),Deilmann_2019}. The unique combination of the properties characteristic of TMD materials, including specific 2D screening \cite{Chernikov:PRL113(2014),Berkelbach:PRB88(2013)}, relatively large reduced mass of an electron-hole pair \cite{Finland2015,Kormanyos2015,Mostaani2017}, nontrivial valley dynamics \cite{Yao2008,Mueller2018,Dufferwiel:NatPhoto11(2017)}, and high binding energies making bright excitons stable even at room temperatures \cite{Ugeda:NatMat:13(2014),Rigosi2016,Wang:RMP90(2018)}\addMisha{,} distinguishes them from conventional semiconductors. Moreover, because of the large binding energy  and small Bohr radius of the excitons, their coupling to light is dramatically increased \cite{Palumno2015,Liu2015,Wang2016,Wang:RMP90(2018)}. Therefore, if a TMD monolayer (ML) is placed inside a photonic cavity, one can relatively easily reach the regime of the strong light-matter coupling, where exciton-polaritons, hybrid half-light, half matter elementary excitations, emerge \cite{Dufferwiel:NatPhoto11(2017),Zhang2018,Schneider2018,Fernandez2019,Lackner2021}. 

However, excitons are not the only type of the bound electron-hole complexes that can be formed. In doped samples, where photocreated excitons interact with a Fermi sea of resident charge carriers, electrons or holes, formation of trions and Fermi polarons \cite{Sidler:NatPhys13(2016),Tan:PRX10(2020)} becomes possible. Similar to the case of excitons, strong light-matter coupling can be routinely achieved for trions/Fermi-polarons resulting in a plethora of mixed light-matter quasi-particles \cite{Shiau_2017,Chang:PRB98(2018),Kyriienko:PRL125(2020),Zhumagulov:NPJ2022, Rana:PRB102(2020),Rana:PRL126(2021),Koksal:PRR3(2021),Ravets:PRL120(2018),Levinsen:PRR1(2019),BastarracheaMagnani:PRL126(2021),LiBleuPRL2021,LiBleuPRB2021,EfimkinPRB2021,Tiene2022,Denning:PRB105(2022),Denning:Arxiv(2021)}.

Besides attaching a free electron, two excitons in a TMD ML can bind together, forming a biexciton, a solid state analog of a hydrogen molecule \cite{You:NatPhys11(2015),Shang2015,Sie2015,Sie2016,Lee2016,Kim2016,Okada2017,Paradisanos2017,Hao2017,Pei2017,Nagler2018,Stevens2018}. The characteristic binding energies can reach 50~meV \cite{You:NatPhys11(2015),Shang2015}, which is more than an order of magnitude higher compared to conventional semiconductor quantum wells, and makes biexcitons in TMD MLs stable up to room temperature. These excitonic molecules underlie basic nonlinear response of 2D semiconductors, reveal intriguing many-body
physics and can serve\addMisha{,} in particular\addMisha{,} as a platform for quantum optics experiments due to their cascaded emission accompanied by entangled photon generation~\cite{Benson2000,Johne2008,He2016}.

Although formation of biexcitons in TMD MLs is well studied both experimentally \cite{You:NatPhys11(2015),Shang2015,Sie2015,Sie2016,Lee2016,Kim2016,Okada2017,Paradisanos2017,Hao2017,Pei2017,Nagler2018,Stevens2018} and theoretically \cite{ivanov98,PhysRevB.61.1692,PhysRevB.62.R7763,Mayers2015,Kylanpaa2015,Szyniszewski2017,Torche2021,PhysRevLett.126.017401,PhysRevB.100.195301,PhysRevB.109.195432}, the biexcitonic effects in TMD-based microcavities and formation of biexciton-based composite half-light half-matter particles was not studied in sufficient detail, to the best of our knowledge. The possibility of forming a bipolariton resonance in conventional semiconductor-quantum-well-based planar microcavities was predicted years ago \cite{Ivanov2004,Wouters2007} and later on experimentally confirmed \cite{Takemura2014}. Further studies mainly concerned conventional semiconductor-quantum-well-based structures focusing on the details of spin-dependent polariton-polariton interactions mediated by biexcitons~\cite{PhysRevB.90.195307,PhysRevB.95.205303}, dephasing~\cite{PhysRevB.94.195301}, photon entanglement in biexciton/bipolariton emission cascades~\cite{PhysRevLett.122.047402}, formation of many-body states~\cite{PhysRevLett.123.266401}, including those arising due to the Feshbach resonances~\cite{PhysRevX.13.031036} and condensate-mediated interexciton interactions~\cite{PhysRevLett.126.017401}. As for the TMD MLs placed in microcavities, Ref.~\cite{PhysRevResearch.2.043185} provides accurate, but perturbative results for the exciton-exciton interactions, while Ref.~\cite{PhysRevB.106.155157} focused on the effects of spin-dependent contributions to the exciton dispersion treating exciton-exciton interaction within a model potential approach. A detailed analysis of the biexciton-photon coupling in TMD materials, in particular, the emerging emission pattern in the regime of strong matter coupling and its microscopic characteristics that would form a solid basis for futher studies of quantum and many-body phenomena mediated by biexcitons/bipolaritons, is still lacking. In this paper, we propose a corresponding microscopic theory. 

The paper is structured in a following way: after a brief introduction (Sec.~\ref{sec:intro}) we formulate the model and derive the parameters of photon interaction with excitons and biexcitons in Sec.~\ref{sec:model}. The results are presented and discussed in Sec.~\ref{sec:res} and the concluding remarks are formulated in Sec.~\ref{sec:concl}. Appendices contain technical details.

\section{Model}\label{sec:model}
\begin{figure}[b!]
    \centering
    \includegraphics[width=\linewidth]{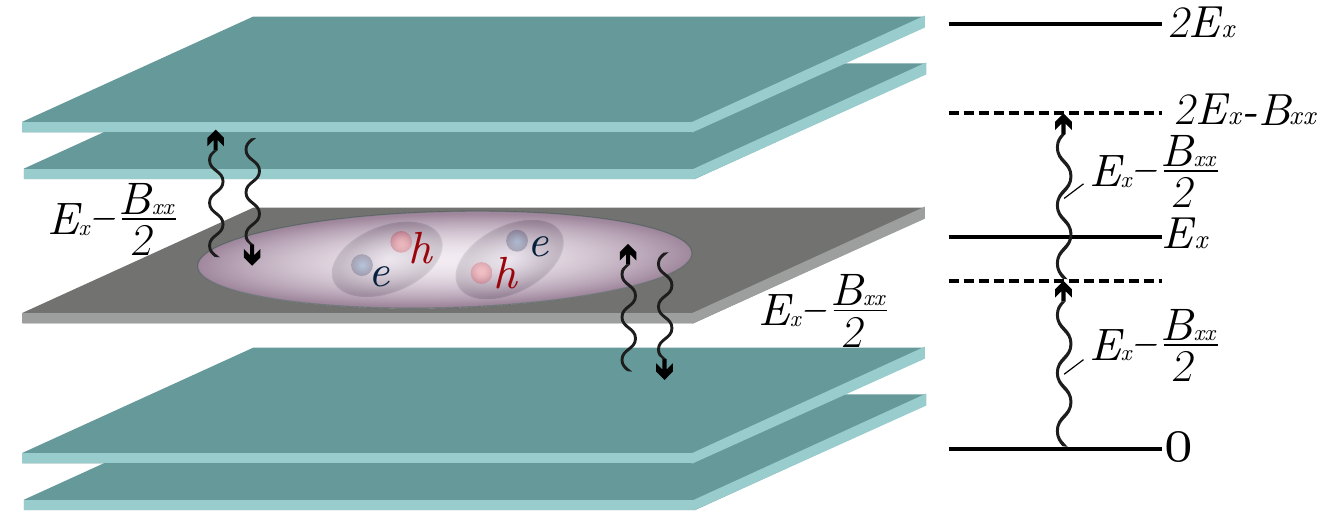}
    \caption{Schematic presentation of the analyzed system and related processes. A monolayer of WS$_2$ is placed between two dielectric Bragg mirrors (DBRs) in an antinode of a confined photonic  cavity mode.
    The cavity mode is resonantly tuned to biexcitonic resonance, so that the photon energy $E_c$ is twice smaller than that of a biexciton $E_b=2E_x-B_{xx}$, where $E_x$ is the exciton energy and $B_{xx}$ is the binding energy of a biexciton. Creation of a biexciton is a two photon process, which makes the behavior of the system strongly dependent on the pump intensity.}
    \label{fig:biexciton_formation}
\end{figure}

We consider the system schematically shown in Fig.~\ref{fig:biexciton_formation}. A monolayer of a TMD material is placed inside a planar microcavity. The confined photonic modes of two opposite circular polarizations are resonantly coupled with excitons in TMD ML. For simplicity, we consider the case corresponding to the in-plane momentum of quasiparticles $\textbf{k}=0$.

Our aim is to present an effective Hamiltonian $\mathcal H$ acting in the basis of the cavity photon, exciton, and biexciton states that describes the light-matter interaction. The parameters of this Hamiltonian are derived within a microscopic model based on the treatment of excitons and biexcitons as two- and four-particle electron-hole complexes, i.e., on fully microscopic grounds. Let $c_\sigma^\dag$ and $c_\sigma$ be the creation and annihilation operators of the cavity mode, $x_\sigma^\dag$ and $x_\sigma$ be the corresponding excitonic operators, and $b^\dag$ and $b$ those of a bright biexciton (which is a singlet both in two electron and two hole spins/valleys) with $\sigma=\pm$ denoting the corresponding circular polarization.  We treat excitons and biexcitons as ideal bosons assuming that their densities $\langle x^\dag_\sigma x_\sigma\rangle/S$, $\langle b^\dag b\rangle/S$ are much smaller than squared Bohr radius ($S$ is the normalization area)~\cite{Combescot2007}. The corresponding model Hamiltonian reads
\begin{multline}
\label{H}
\mathcal H = \sum_{\sigma=\pm}E_c c_\sigma^\dag c_\sigma + \sum_{\sigma=\pm} E_x x_\sigma^\dag x_\sigma + E_b b^\dag b  \\ +\sum_{\sigma=\pm} V_x(c_\sigma^\dag x_\sigma + x_\sigma^\dag c_\sigma) \\
+ V_b(c_+^\dag x_-^\dag b + b^\dag x_- c_+ + c_-^\dag x_+^\dag b + b^\dag x_+ c_-).
\end{multline}
Here $E_c$ is the cavity mode energy, $E_x$ is the exciton energy, $E_b$ is the biexciton energy, $V_x$ and $V_b$ are the exciton- and biexciton-light coupling constants to be found below from microscopic model. Two first terms and the fourth term represent the conventional exciton-photon interaction Hamiltonian in a microcavity~\cite{microcavities}. We assumed that the main process of the biexciton formation is the cascade process of $c\to x\to b$. The process of the direct two-photon absorption and creation of biexciton via remote states  is weaker and can be analyzed separately \cite{Pervishko2013}. We assume that the differences $E_c - E_x$ and $E_b - 2E_x$ are by far smaller than the energies of other states making it possible to keep only the processes described by the Hamiltonian~\eqref{H}. Moreover, we examine only the processes where zero momentum dominates and scattering into excitons with a momentum different from zero introduces minor changes mainly providing an additional relaxation channel for biexcitons without qualitatively affecting the observed picture. We also consider the case of an undoped microcavity, and so neglect the possibility of the formation of trions and exciton polarons. In this work we disregard the exciton-phonon interaction that gives rise a number of phonon replicas in optical emission spectra~\cite{PhysRevLett.132.036903}. The oscillator strengths of these replicas in absorption are negligibly small, so the role of exciton-phonon coupling is expected to diminish in microcavities. It follows from the symmetry that biexcitons have angular momentum zero and are formed from the opposite spin (circular polarization) excitons. Under such conditions, we can neglect energy blueshifts due to the same spin excitons.

It is important to note that the combination of strong spin-orbit interaction and violation of the inversion symmetry of the system leads to additional valley degrees of freedom. The valley structure leads to the fact that the optical transition  being direct is sensitive to the choice of light polarization and can only be induced in the so-called $K_+$ or $K_-$ valleys.

To determine the parameters $V_x$ and $V_b$ the microscopic approach is needed and both exciton and biexciton states should be decomposed into the states of electrons and holes. To that end, we recall that the light-matter interaction in the basis of electrons and holes can be described by the following Hamiltonian
\begin{equation}
\label{light-matter}
H_{\textup{l-m}} = \sum_{\bm k,\sigma=\pm} \Omega (a^\dag_{e,\bm k,\sigma} a^\dag_{h,-\bm k,\sigma}c_\sigma +{\rm h.c.}), 
\end{equation}
where the constant $\Omega$ is proportional to the electric field in the mode and interband dipole matrix element, $a^{\dag}_{e,\bm k,\sigma}$, $a^\dag_{h,\bm k,\sigma}$ are the electron and hole creation operators (correspondingly, $a_{e,\sigma k}$, $a_{h,\sigma k}$ are the electron and hole annihilation operators). 
Note also that for charge carriers $+$ and $-$ denote spin  components (or valley indices) and we assume that the optical transition selection rules correspond to the simplest case $$\sigma=+ \quad \Rightarrow\quad (e+, h+); \quad \sigma=- \quad \Rightarrow \quad (e-, h-).$$ In Eq.~\eqref{light-matter} we neglected the wave vector of light so that the momenta of the photoexcited electron and hole are opposite. 
 
The exciton state can be represented as~\footnote{We use the envelope-function approach, where the wavefunction is represented as a product of the Bloch function and a smooth envelope. The latter depends only on the in-plane coordinates/momenta of the quasiparticles. The Bloch amplitudes determine the quantity $\Omega$ in Eq.~\eqref{light-matter} via the interband momentum matrix element $d_{\rm cv}$ or `interband' velocity, see Eq.~\eqref{eqn:v_x}.}:
\begin{equation}
\label{exc}
\Psi_{x,\sigma} = \sum_{\bm k_e,\bm k_h} F({\bf k}_e,{\bf k}_h) a^\dag_{e,{\bf k}_e,\sigma} a^\dag_{h,{\bf k}_h,\sigma}|0\rangle,
\end{equation}
where $|0\rangle$ is the ground state of the system with empty conduction band and filled valence band and $F(\bm k_e,{\bf k}_h)$ is the Fourier transform of a two-particle envelope function,
\begin{equation}
F({\bf k}_e, {\bf k}_h) = \dfrac{1}{S}\int {\rm d}{\bf r}_e {\rm d}{\bf r}_h \psi_{{\bf K}_{\textup{ex}}}({\bf r}_e, {\bf r}_h) e^{-\mathrm i {\bf k}_e {\bf r}_e - \mathrm i {\bf k}_h {\bf r}_h}.
\end{equation}
As already noted, in this paper, we restrict ourselves to consideration of a biexciton consisting of excitons with a zero wave vector. In this case, $\bm K_{\textup{ex}} \equiv \bm k_e + \bm k_h =0$ and $F({\bf k}_e,{\bf k}_h)\propto \delta_{{\bf k}_e, -{\bf k}_h}$. Combining Eqs.~\eqref{light-matter} and \eqref{exc}, we obtain:
\begin{multline}
\label{Vx}
V_x = \langle \Psi_{x,\sigma} | H_{\textup{l-m}}c_\sigma^\dag|0\rangle = \Omega \sum_{\bf k} F^*({\bf k}, - {\bf k})=\\
=\Omega\int {\rm d}{\bf r} \psi_{{\bf K}_{\textup{ex}}=0}^*({\bf r}, {\bf r}) =\sqrt{S} \Omega \varphi_x^*(0),
\end{multline}
where $\varphi_x(\rho)$ is the envelope function of the relative motion of the electron and hole in the exciton
\begin{align}
\psi_{{\bf K}_{\textup{ex}}}({\bf r}_e, {\bf r}_h) =\dfrac{1}{\sqrt{S}}\,\varphi_x(|{\bf r}_e - {\bf r}_h|) e^{\mathrm i {\bf K}_{\textup{ex}} {\bf R}}, \label{eqn:full_wf}
\end{align}
where ${\bf R} = (m_e {\bf} {\bf r}_e+ m_h {\bf r}_h)/M$, and the total mass of the exciton is defined as  $M=m_e+m_h$. The parameter $V_b$ can be evaluated in a similar way. Representing the biexciton wave function via the electron and hole creation and annihilation operators and the biexciton envelope function $\psi_b$ and taking into account its antisymmetry with respect to permutations of identical fermions we obtain [see Appendix~\ref{appendix:interactionV} for details]
\begin{align}
\label{Vb}
    V_b=\Omega\!\int \!\! {\rm d}{\bf r}_1{\rm d}{\bf r}_2 \varphi_{x}(|{\bf r}_2|) \psi({\bf r}_1-{\bf r}_2,{\bf r}_1,{\bf r}_2/2).
\end{align}
Here we made use of the fact that the biexciton envelope function $\psi_b$ can be represented as a product of the relative motion envelope and the exponent responsible for the motion of the biexciton as a whole:
\begin{align}
\label{eq:bie_wf}
\!\!\psi_b({\bf r}_e,\! {\bf r}_e',\!{\bf r}_h,\!{\bf r}_h') = \frac{1}{\sqrt{S}} \, \psi({\bf r},{\bf r}',{\bf R}) e^{i{\bf K}_b{\bf R}_b},
\end{align}
where $\bm K_b$ is the biexciton translational motion wave vector, ${\bf r}={\bf r}'_e-{\bf r}_e$, ${\bf r}'={\bf r}'_h-{\bf r}_h$,  ${\bf R}=({\bf r}_e+{\bf r}'_e-{\bf r}_h-{\bf r}'_h)/2$, ${\bf R}_b=[m_e({\bf r}_e+{\bf r}'_e)+m_h({\bf r}_h+{\bf r}'_h)]/2M$. 

The function $\psi$ in Eq.~\eqref{eq:bie_wf} can be found variationally extending the approach developed in Ref.~\cite{Kleinman1983} for TMD ML case where the Coulomb interaction is described by the Rytova-Keldysh potential~\cite{Rytova1967,Keldysh1979}. The trial function is chosen in the form
\begin{multline}
  \psi({\bf r},{\bf r}',{\bf R}) = \dfrac{\left[ (kv)^{\nu}\!\exp{\left(-\rho k v\right)}+\lambda\exp{\left(-\tau k v \right)}\right]}{\sqrt{N_b}}\\ 
    \times\exp{\left[-\dfrac{k(s_1+s_2)}{2}\right]}\cosh{\left[\dfrac{\beta k (t_1-t_2)}{2}\right]},\label{eqn:phi_b_normalized}
\end{multline}
where $s_1 = r_{eh} + r_{eh'}$, $t_1 = r_{eh} - r_{eh'}$, $s_2 = r_{e'h} + r_{e'h'}$, $t_2 = r_{e'h} - r_{e'h'}$, $u = r_{ee'}$, $v = r_{hh'}$, $r_{ij} \equiv |{\bf r}_i-{\bf r}_j|$, and $N_b$ is a normalization factor (see Appendix~\ref{appendix:wavefunctions} for details). The variational parameters are the following: dimensional $k$ and dimensionless $\beta$, $\nu$, $\rho$, $\lambda$, and $\tau$. They are found by maximizing the total binding energy $B_{2e,2h}$ of the four-particle system. Using the trial function~\eqref{eqn:phi_b_normalized}, we obtain for the exciton-biexciton conversion matrix element:
\begin{equation}
V_b=\alpha \frac{a_{\textup{B}}}{\sqrt{S}}V_x,\label{Vb}
\end{equation}
where dimensionless parameter $\alpha$ reads:
\begin{multline}
\label{Eq:alpha}
    \alpha=\dfrac{2\pi}{\tilde{k}\sqrt{\tilde{N}_b}}\int\limits_0^{\infty}x{\rm d}x\int\limits_0^{\infty}y{\rm d}y\int\limits_0^{2\pi}{\rm d}\theta \ \phi{\left(y/\tilde{k}\right)}\\
    \times \exp{\left(-\dfrac{x+y+\sqrt{x^2+y^2-2xy\cos{\theta}}}{2}\right)}\\
    \times \cosh{\left[\dfrac{\beta\left(y-x-\sqrt{x^2+y^2-2xy\cos{\theta}}\right)}{2}\right]}\\
    \times \left[x^{\nu}\exp{\left(-\rho x\right)}+\lambda\exp{\left(-\tau x\right)}\right].
\end{multline}
Here we have introduced dimensionless parameter $\tilde{k}=a_{\textup{B}}k$ and normalization  factor $\tilde{N}_b = a_{\textup{B}} N_b$, where $a_\textup{B}$ is the Bohr radius. Note that the relevant center-of-mass momenta of the quasiparticles are, in any case, of the order of light wavevector and neglibigle compared to the inverse Bohr radii of excitons and biexcitons. Thus, the momentum dependence of $V_x$ and $V_b$ {in Eq.~\eqref{Vb} can be disregarded.

The Hamiltonian~\eqref{H} allows us to formulate the set of the equations of motion for the system. We first write the Heisenberg equations of motion for the operators $c_\pm,x_\pm,b$ and then use the mean-filed approximation, replacing the second-quantization operators by corresponding classical fields, $c_\pm\rightarrow C_\pm =\langle c_\pm\rangle$, and corresponding amplitudes $x_\pm\rightarrow X_\pm=\langle x_\pm \rangle$ and $b\rightarrow B=\langle b\rangle$ where the angle brackets $\langle \ldots \rangle$ denote a quantum mechanical averages. The resulting system reads [cf. Ref.~\cite{ivchenko2005optical}, Eqs. (7.11)--(7.13)]
\begin{subequations}
\label{motion}
\begin{align}
\mathrm i \hbar\frac{{\rm d}C_\pm}{{\rm d}t} =& E_c C_\pm + V_x X_\pm + V_b X_\mp^* B ,\label{eqn:sys_eq_12}\\
\mathrm i \hbar\frac{{\rm d} X_\pm}{{\rm d}t} =& E_x X_\pm + V_x C_\pm + V_b C_\mp^* B,\label{eqn:sys_eq_34}\\
\mathrm i \hbar\frac{{\rm d} B}{{\rm d}t} =& E_b B + V_b (X_-C_+ + X_+ C_-).\label{eqn:sys_eq_5}
\end{align}
\end{subequations}
Solution of Eqs.~\eqref{motion} with appropriate excitation conditions allows one to describe the response of a microcavity with allowance for the excitons and biexcitons formation. It is presented and discussed in the next Sec.~\ref{sec:res}.

\section{Results and discussion}\label{sec:res}

In this paper we consider the case of a WS$_2$ ML placed in a microcavity (the cases of the other TMD materials can be analyzed similarly).  We use the following set of material parameters: the band gap $E_g=2.238$~eV~\cite{Goryca2019}, effective dielectric constant $\varepsilon$ and screening radius $r_0$ in the Rytova-Keldysh potential are chosen to be $4.4$ and $0.89~\textup{nm}$, respectively~\cite{Goryca2019,Kormanyos_2015,PhysRevB.96.115409,Chernikov:PRL113(2014)}. The effective masses of the carriers are $m_e=0.26 m_0$ and $m_h=0.35 m_0$~\cite{Kormanyos_2015}, which leads to the electron-hole reduced mass $\mu=m_em_h/M=0.149 m_0$ and electron-hole mass ratio $\sigma=m_e/m_h=0.743$. By numerically solving the Schr\"{o}dinger equation for selected values of the parameters (see Appendix~\ref{appendix:wavefunctions}), we found the following value of the exciton binding energy: $B_{x}=152~\textup{meV}$. Thus, the energy of the exciton transition is $E_{x}=E_g - B_x = 2.086~\textup{eV}$ that roughly corresponds to the experimental data~\cite{zipfel2019exciton}.

For the light-matter coupling constant $\Omega$ determining $V_x$ via Eq.~\eqref{Vx}, we use the standard expression (using the SI units): $\Omega=\sqrt{E_{c}/\varepsilon_b\varepsilon_0 V}d_{\textup{cv}}$, where $V=LS$ is the cavity volume, $L=\pi c \hbar/ n_b E_{c}$ is the effective cavity length, $n_b$ is the effective refractive index for the area between the mirrors, for which we chose a moderate value of $1.4$, and $d_{\rm cv}$ is the interband dipole matrix element. The latter is expressed within the $\mathbf k \mathbf p$-model as $d_{\textup{cv}}=e\hbar v_{F}/E_{g}$, where the `interband' velocity $v_F$ for WS$_2$ amounts to $6.65\times 10^5$ m/s~\cite{HUONG2020114315,PhysRevLett.108.196802}. 

Finally, we present the exciton envelope function at coinciding coordinates $\varphi_x(0)= \eta/a_{\textup{B}}$, where the Bohr radius is defined as  $a_{\textup{B}}=4\pi\varepsilon_0\varepsilon\hbar^2/e^2\mu$ and the dimensionless parameter $\eta=0.615$ is extracted from the numerical ground-state solution of the excitonic Schr\"{o}dinger equation.
Taking into account all of the above, we obtain
\begin{equation}\label{eqn:v_x}
V_x=\dfrac{\eta v_F e \hbar}{E_g a_{\textup{B}} }\sqrt{\dfrac{E_{c}}{\varepsilon_b \varepsilon_0 L}}.
\end{equation}
 Using the set of the parameters specified above and $E_c=2.0815$~eV, we get $V_x=23.14~\textup{meV}$, thereby for the Rabi splitting $2V_x$ we have $\sim 46.3$~meV. This value is quite comparable with those obtained previously both in theoretical and experimental works on TMDs in the range of $20-70$~meV~\cite{Barachati2018,PhysRevResearch.3.L032056,Flatten2016}.

Maximization of the biexcitonic binding energy with variational biexcitonic wavefunction \eqref{eqn:phi_b_normalized} allows us to determine the biexciton binding energy $B_{xx} \approx  8.93~\textup{meV}$ and the parameter $\alpha = 1.016$, that determines the biexciton to light coupling constant $V_b$ in Eq.~\eqref{Vb}. We assume that the cavity is tuned to the biexciton resonance, hence $2E_c=E_b \equiv 2E_x - B_{xx}$.
Note that unlike $V_x$, $V_b$ explicitly depends on the sample area $S$. To get rid of this dependence, which does not influence the systems dynamics due to the $S$-dependence of the  amplitudes, one can pass in Eqs.~\eqref{motion} to the normalized  amplitudes, $C_{\pm} \rightarrow \tilde{C}_{\pm} \sqrt{S}$ [Appendix~\ref{appendix:dimensionless}], and then solve numerically the resulting system which does not contain the normalization factor anymore. 

\subsection{Transient response}\label{subsec:transient}

First, let us focus on the transient response of the microcavity after a short optical pump pulse that creates non-zero photonic occupancy $C_\pm(t)\neq0$ at the initial time instant $t=0$, analyzing the dynamics in real time and get the corresponding energy spectrum via its Fourier transform.  The details on the steady state dynamics are presented below in Sec.~\ref{subsec:steady}.

\begin{figure}
    \centering
    \includegraphics[width = \linewidth]{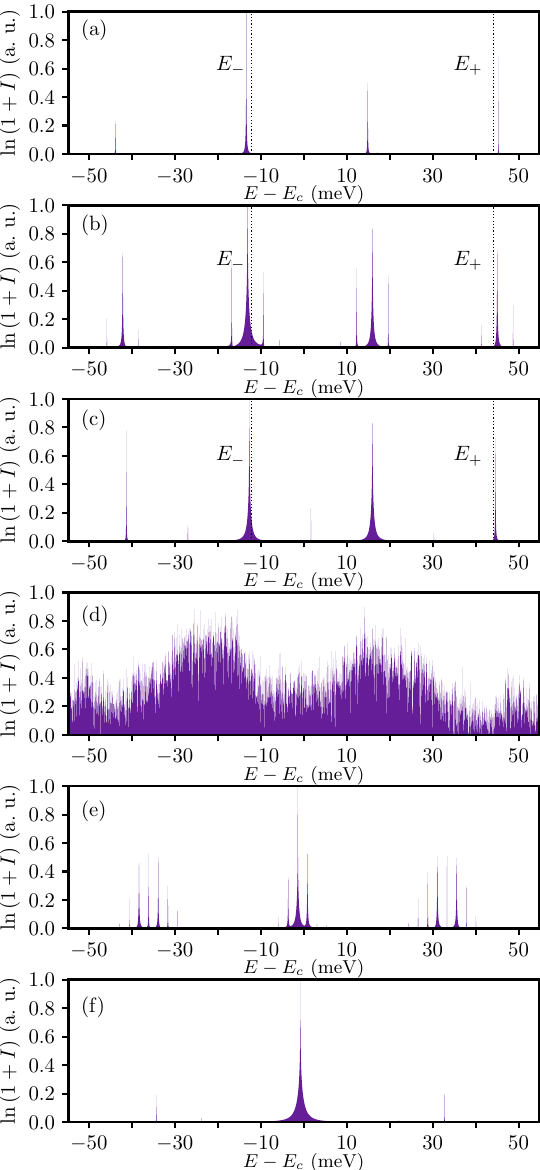}
    \caption{Emission spectra of the system for various values of pumping magnitude characterized by total number of the excitations $N$. The  upper panels (a), (b), and (c) correspond to relatively weak pumps, where two main peaks resembling Rabi-like splitting, together with a third weak peak are observed. The panel (d) illustrates a transient regime of intermediate pumps without regular structure of the peaks, characteristic to the chaotization of the nonlinear dynamics of the system. The lower panels (e) and (f) demonstrates the formation of clear Mollow-type triplet pattern which is observed for high pumps.}
    \label{fig:spectra}
\end{figure}

To get the emission spectrum, one needs to calculate a Fourier transform:\footnote{In the absence of damping the integrals diverge at $t\to \infty$, hence, a small damping is added to obtain the numerically converging results.}
\begin{equation}
    \label{I:spec}
    I(E) =\left|\int \limits_{0}^\infty  \tilde{C}(t)  e^{\mathrm i E t} dt \right|^2.
\end{equation}
The corresponding spectra are plotted in Fig.~\ref{fig:spectra} for the six different pumps, corresponding to the initial number of photons in a cavity $N$ defined as $\tilde{C}_{\pm}(\tilde{t}=0)= \sqrt{N}$, $\tilde{X}_{\pm}(\tilde{t}=0)=0$, and $\tilde{B}(t=0) = 0$. The chosen cases illustrate the key regimes characterizing biexciton-polariton formation: (a), (b), and (c) small to moderate $N < N_{c,1}$, (d) intermediate $N_{c,1} < N < N_{c,2}$, and (e) and (f) large $N>N_{c,2}$.

In the regime (a), when $N \ll N_{c,1}$, two main peaks resembling a Rabi doublet are seen in the spectrum (corresponding to $E-E_c\approx\pm 12$~meV), together with an additional peak located by about $45$ meV above $E_c$, see Fig.\ref{fig:spectra}a. These three peaks originate from the mixture of photonic, excitonic and biexcitonic modes. If one switches off the photon-biexciton coupling putting $\alpha=0$ or considers negligibly small exciton occupancy, the central peak disappears, and flanking peaks continuously approach the positions of upper and lower exciton-polaritons, shown in the figure by vertical dashed lines and labeled by $E_{\pm}$. Note, that as the formation of a biexciton is a two-photon process, the distance between the two main peaks increases with the photon occupancy~\cite{Pervishko2013}, that we directly observed for small pump values.
It is interesting to note also that an increase in pump at some point causes subpeaks to appear [panel (b) in Fig.~\ref{fig:spectra}], which are localized in the vicinity of the main ones; immediately before the onset of the transitional chaotic regime, the subpeaks disappear [panel (c) in Fig.~\ref{fig:spectra}].

If the pump exceeds some critical value, the above described regular peak structure disappears, and the spectrum is characterized by a broad distribution of multiple peaks characteristic to the transition to chaotic dynamics of the system described by nonlinear set of equations \eqref{eqn:sys_eq_12}--\eqref{eqn:sys_eq_5}, see Fig.~\ref{fig:spectra}(d). This intermediate regime is related to the multistability behavior of the polariton system akin studied in Refs.~\cite{Gippius:2005eng,PhysRevLett.98.236401,Gavrilov:2020eng}.

Finally, for sufficiently high pumps the spectrum again becomes regular, but now the pattern closely resembling the Mollow triplet is formed, see panels (e) and (f) in Fig.~\ref{fig:spectra}. The strong central peak is located around the position corresponding to the cavity photon, while the distance between the satellite peaks increases with increase of the pump. Note that after the transition from the chaotic regime, the main peaks are accompanied by additional satellite resonances [Fig.~\ref{fig:spectra}(e)], which completely disappear with increasing pumping [Fig.~\ref{fig:spectra}(f)]. This behavior is very similar to what is happening in the paragon case of Jaynes-Cummings model. However, differently from this latter case, the Rabi-doublet and Mollow-triplet regimes are separated by the chaotic regime with broad spectral distribution. We note that a spectrum similar to panel~(e) was already shown earlier in the analysis of the quantum dot model~\cite{PhysRevB.93.115308}.

\AK{\begin{figure*}[t] 
\centering
\includegraphics[width=\textwidth]{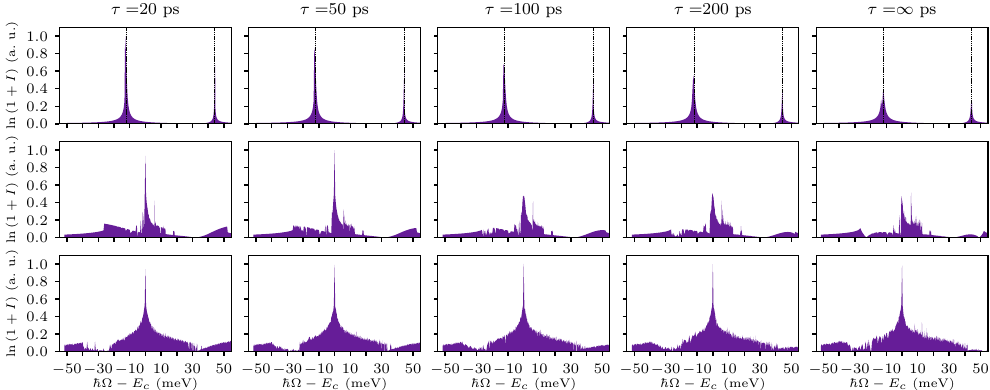} 
\caption{Nonlinear reflectance spectra of the system for various values of the pumping magnitude and lifetime $\tau$/ damping (different pumping corresponds to different rows ($N=10^{-9},\, 9\cdot 10^{-6},$ and $10^{-4}$ for upper, center, and bottom rows, respectively), while the damping value changes along the columns).}
\label{fig:damping}
\end{figure*}}

\subsection{Steady-state response}\label{subsec:steady}

Let us now turn to the steady state excitation regime.
In this section, we present an example of the analysis of the ``nonlinear'' reflection spectrum when a constant pump is applied to the system, its frequency is scanned across the relevant range in the vicinity of a cavity mode, and the reflected intensity is calculated. To analyze such a situation, we explicitly include the pumping and the decay terms in the system~\eqref{motion}:
\begin{subequations}
\label{motion_2}
\begin{align}
\mathrm i \frac{{\rm d}\tilde{C}_\pm}{{\rm d}\tilde{t}} =& \tilde{e}_c \tilde{C}_\pm + v_x \tilde{X}_\pm + \alpha v_x \tilde{X}_\mp^* \tilde{B} + \sqrt{N}e^{-\mathrm i\omega\tilde{t}} ,\label{si_eqn_2:sys_eq_12}\\
\mathrm i \frac{{\rm d} \tilde{X}_\pm}{{\rm d}\tilde{t}} =& \tilde{e}_x \tilde{X}_\pm + v_x \tilde{C}_\pm + \alpha v_x  \tilde{C}_\mp^* \tilde{B},\label{si_eqn_2:sys_eq_34}\\
\mathrm i \frac{{\rm d} \tilde{B}}{{\rm d}\tilde{t}} =& \tilde{e}_b \tilde{B} + \alpha  v_x (\tilde{X}_-\tilde{C}_+ + \tilde{X}_+ \tilde{C}_-),\label{si_eqn_2:sys_eq_5}
\end{align}
\end{subequations}
where $\omega=\hbar\Omega/E_c$ is a dimensionless frequency of an external pump and $\tilde{e}_i=e_i-\mathrm i \delta$ are dimensionless shifted energies taking into account the damping in the system. Here we define  $\delta=\hbar/(2\tau E_c)$, where $\tau$ is the effective lifetime. Here, for simplicity, we do not distinguish between the lifetimes of excitonic complexes and cavity mode, the extension of Eqs.~\eqref{motion_2} for the general case is rather trivial. The spectrum was constructed in a similar way as described in Sec.~\ref{subsec:transient}, however, here we changed the frequency of the external pump $\Omega$ and analyzed the value of the squared absolute value of the Fourier amplitude precisely at a frequency equal to that of the pump, $E=\hbar\Omega$. The results are plotted in Fig.~\ref{fig:damping} where we also analyze the effect of damping on the reflectance spectrum. To this end, we consider the spectra for different values of $\tau$. For simplicity, we have chosen this value to be the same for all amplitudes (for all equations) since our main goal is to demonstrate the fact that the chaotic regime remains in any case. In Fig.~\ref{fig:damping}, we show the damping effect for the following lifetimes: $20$, $50$, $100$, $200$ and $\infty$ picoseconds.

At weak excitation (top row in Fig.~\ref{fig:damping}) the spectrum contains to peaks related to the exciton-polariton models, similarly to the linear regime shown in the  panel (a) in Fig.~\ref{fig:spectra}. The natural linewidth of sharp peaks is $\hbar/(2\tau)$ and for all relevant values of the lifetimes $\tau$ is sufficiently small, so the peaks are extremely sharp in the center. The wings around the peaks are due to the nonlinear effects and are, actually, weak, note the logarithmic scale of the intensity. 

An increase in the pump intensity, naturally, enhances the role of biexciton nonlinearity and brings system to a chaotic-like regime, compare middle row in Fig.~\ref{fig:damping} and Fig.~\ref{fig:spectra}(d). Weak wings of the peaks transform to an uneven background reflecting multistable behavior of the system. Further increase in the pumping (bottom row in Fig.~\ref{fig:damping} results in formation of sharp peak at $\hbar\Omega = E_c$ on top of the noisy background. In this regime, in fact, the strong coupling is lost, cf. Fig.~\ref{fig:spectra}(f). Hence, we conclude that the chaotic regime can be realized in the presence of damping, while the Mollow-triplet like regime, clearly observed in the transient case at $\tau\to \infty$ [Fig.~\ref{fig:spectra}(d)] turns out to be more fragile. 

To probe the nonlinear effects described here, one could perform angle-resolved reflectance or photoluminescence (PL) measurements while varying the incident laser intensity. At low pump power, a conventional Rabi-like doublet would be visible in reflectance (or absorption) and the PL spectra. By gradually increasing the excitation strength, one can detect the onset of an irregular, multi-peak emission corresponding to the chaotic regime. Finally, at higher powers, a single peak should reappear indicating the loss of the strong coupling regime. These transitions can also be characterized by time-resolved spectroscopy, where ultrafast pulses reveal transient oscillations in the emission intensity. Analyzing these traces in the time domain could further confirm the chaotic dynamics, for example, by inspecting the sensitivity with respect to the changes of the initial conditions.

To understand why the spectrum has this structure, we can proceed to the search for solutions at the external pump frequency. Then we will obtain a system of nonlinear algebraic equations, which can already be processed both numerically and using perturbation theory, presenting all solutions in the form of expansion in $\alpha$: $\tilde{C}_{\pm}=c_0 + c_1\alpha + c_2\alpha^2+\dots$. Having done all this, one can reveal an abundant number of solutions, leading to multi-stability. In particular, the drawdown in the center of each of the two peaks in Fig.~\ref{fig:damping}a at a low pump level is already visible when searching for solutions up to $\alpha^2$.


\section{Conclusion}\label{sec:concl}
We developed a microscopic theoretical approach for studying the emission spectrum of a quantum microcavity with an embedded TMD ML accounting for the resonant coupling of a cavity mode with both excitonic and biexcitonic transitions. We have presented an effective Hamiltonian describing the light-matter interaction in microcavities with allowance for the biexciton formation. Further, we have adapted a special type of variational ansatz which takes into account the symmetry properties of the biexciton in the case of transition metal dichalcogenides to find the biexciton wave function, binding energy, as well as the exciton-biexciton conversion matrix element. As creation of a biexciton requires absorption of two photons with opposite circular polarizations, the formation of a biexciton-polariton is a strongly nonlinear process. It was first shown that for weak pumps it is characterized by a Rabi-type doublet, while for large pumps radical reshaping of the emission spectrum takes place and the pattern resembling the Mollow triplet at a weak damping and single peak at a moderate to strong dampings emerges. This two regimes are separated by an intermediate regime characterized by broad multi-peak emission characteristic to chaotic dynamics of the system. The latter regime can be particularly interesting in view of emergent interest to chaotic and regular nonlinear behavior in solid-state systems related to the search of the time-crystalline behavior~\cite{Carraro-Haddad:2024aa,Greilich:2024aa}.

While we focused here on the microcavity structures with embedded TMD monolayers like experimentally studied in Refs.~\cite{Flatten2016,Dufferwiel:NatPhoto11(2017),Barachati2018}, similar effects are, in principle, possible for other TMD-based structures like in superlattices with alternating TMD and hBN layers~\cite{PhysRevB.108.L121402}.

\begin{acknowledgements}
The study is supported by the Ministry of Science and Higher Education of the Russian Federation (Goszadaniye), project No. FSMG-2023-0011. The work of A.K. is supported by the Icelandic Research Fund (Ranns\'oknasj\'oður, Grants No. 2410550).  
\end{acknowledgements}

\appendix
\allowdisplaybreaks

\section{Interaction constants}\label{appendix:interactionV}
As was shown in the main text of the paper, the interaction constant $V_b$ can be found by means of the following equation:
\begin{multline}
\label{Vb}
V_b =  \langle \Psi_b |\mathcal H_{\textup{l-m}} c_-^\dag| \Psi_{x,+}\rangle =\Omega \int {\rm d}{\bf r} {\rm d}{\bf r}_e {\rm d}{\bf r}_h
\\ \times \psi_b^*({\bf r}_e, {\bf r}, {\bf r}_h, {\bf r})\psi_{{\bm K}_{\textup{ex}}}({\bf r}_e, {\bf r}_h),
\end{multline}
where we assume the following form of the biexciton wavefunction:
\begin{multline}\label{biexc}
\!\Psi_{b}\! =\! \sum\limits_{\substack{\bm k_e,\bm k_h \\ \bm k_e',\bm k_h'}} B({\bf k}_e,{\bf k}_e',{\bf k}_h,{\bf k}_h') \\
\times\frac{a^\dag_{e,\bm k_e,+} a^\dag_{e,\bm k_e',-}-a^\dag_{e,\bm k_e,-} a^\dag_{e,\bm k_e',+}}{\sqrt{2}}\\
\times \frac{a^\dag_{h,\bm k_h,+} a^\dag_{h,\bm k_h',-}-a^\dag_{h,\bm k_h,-} a^\dag_{h,\bm k_h',+}}{\sqrt{2}}|0\rangle,
\end{multline}
in which we have explicitly antisymmetrized the two-electron and two-hole functions. The coefficients $B({\bf k}_e,{\bf k}_e',{\bf k}_h,{\bf k}_h')$ can be found via envelope biexciton function by means the following relation:
\begin{multline}
B({\bf k}_e,{\bf k}_e',{\bf k}_h,{\bf k}_h')=\dfrac{1}{S^2} \int {\rm d}{\bf r}_e {\rm d}{\bf r}_h {\rm d} {\bf r}_e' {\rm d}{\bf r}_h' 
\\ \times \psi_b({\bf r}_e, {\bf r}_e',{\bf r}_h,{\bf r}_h') e^{-\mathrm i {\bf k}_e {\bf r}_e - \mathrm i {\bf k}_h {\bf r}_h-\mathrm i {\bf k}_e' {\bf r}_e' - \mathrm i {\bf k}_h' {\bf r}_h'}.
\end{multline}
If  we   assume the fact that biexciton envelope function $\psi_b$ can be represented as
\begin{align}
\label{eq:bie_wf:appendix}
\!\!\psi_b({\bf r}_e,\! {\bf r}_e',\!{\bf r}_h,\!{\bf r}_h')\!=\!\dfrac{e^{i{\bf K}_b{\bf R}_b}}{\sqrt{S}}\psi({\bf r},{\bf r}',{\bf R}),
\end{align}
then Eq.~\eqref{Vb} for $V_b$ can be rewritten as follows
\begin{align}\label{Vb1}
   V_b=\Omega\!\int \!\! {\rm d}{\bf r}_1{\rm d}{\bf r}_2 \varphi_{x}(|{\bf r}_2|) \psi({\bf r}_1-{\bf r}_2,{\bf r}_1,{\bf r}_2/2).
\end{align}
Thus, to calculate the interaction constant, it is necessary to determine two wave functions $\varphi_{x}(|{\bf r}_2|)$ and $\psi({\bf r}_1-{\bf r}_2,{\bf r}_1,{\bf r}_2/2)$. In the next two sections, we obtain the exciton and biexciton wave functions.

\section{Exciton and biexiton binding energies and wave functions}\label{appendix:wavefunctions}
\subsection{Exciton}
The Hamiltonian of the electron-hole pair moving within the TMD quantum well for a zero center-of-mass momentum reads
\begin{align}\label{eqn:Hamiltoian_dim_ex}
    H_x=-\dfrac{\hbar^2}{2\mu}\Delta-V_{eh},
\end{align}
where the Coulomb interaction is given by the Rytova-Keldysh potential~\cite{Rytova1967,Keldysh1979}:
\begin{align}\label{eq:K-R_potential}
    V_{ij}=\dfrac{e^2}{4\pi\varepsilon_0\epsilon}\dfrac{\pi}{2 r_0}\left[H_0\left(\dfrac{r_{ij}}{r_0}\right)-Y_0\left(\dfrac{r_{ij}}{r_0}\right)\right].
\end{align}
Here $H_0$ and $Y_0$ are the zero-order Struve function and zero-order Bessel function of the second kind, respectively. This modification of the conventional Coulomb potential is dictated by the finite atomic thickness of the sample and its dielectric properties characterized by effective screening length $r_0$ and average dielectric constant $\varepsilon$.  For convenience, let us switch to dimensionless units, where energies and distances are measured in units of Rydberg constant $\mathcal{R}$ and Bohr radius $a_{\textup{B}}$, respectively:
\begin{align}\label{eqn:bohr_rydeberg}
    a_{\textup{B}}=\dfrac{4\pi\varepsilon_0\varepsilon\hbar^2}{e^2\mu}, \quad \mathcal{R}=\dfrac{e^2}{8\pi\varepsilon_0\varepsilon a_{\textup{B}}}.
\end{align}
The dimensionless potential $\tilde{V}_{ij}$ has the following form:
\begin{align}
    \tilde{V}_{ij}\equiv \tilde{V}(x_{ij})=\dfrac{\pi}{2 x_{0}}\left[H_0\left(\dfrac{x_{ij}}{x_{0}}\right)-Y_0\left(\dfrac{x_{ij}}{x_{0}}\right)\right],
\end{align}
where $x_{ij}=r_{ij}/a_{\textup{B}}$. In what follows, $x$ will be used as a dimensionless variable measured in units of   $a_{\textup{B}}$. Thus, the dimensionless energy operator can be written as
\begin{align}\label{eqn:Hamiltoian_dimless_ex}
    \tilde{H}_x=-\tilde{\Delta}-2\tilde{V}_{eh}. 
\end{align}
\begin{figure}[t]
    \centering
    \includegraphics[width = 0.9\linewidth]{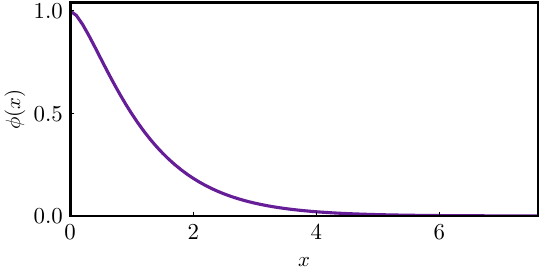}
    \caption{Behavior of the normalized exciton
    wavefunction  $\phi(x)$ constructed by numerically solving the Schr\"{o}dinger equation with the Hamiltonian~\eqref{eqn:Hamiltoian_dimless_ex}.}
    \label{fig:app_ex_wf}
\end{figure}

\noindent The spectrum is a function of the dimensionless parameter $x_0=r_0/a_{\textup{B}}$ depending on the choice of material. As stated in the body of the paper, we consider the WS$_2$ monolayer. In this case,  Bohr radius $a_{\textup{B}}=1.56~\textup{nm}$, which leads to $x_0=0.57$. The numerical solution of this equation yilds the following value for the dimensionless ground-state (GS) energy: $\tilde{E}_{\textup{GS}}=-1.450(1)$. Taking into account the value of the Rydberg constant ($\mathcal{R}=105~\textup{meV}$),  we obtain the exciton binding energy $B_x=152~\textup{meV}$ defined as
\begin{align}
 B_x=-\int{\rm d}{{\bf r}}\varphi_x(|{{\bf r}}|)^* H_x \varphi_x(|{\bf r}|).   
\end{align}
The GS solution $\phi(x)$ of the Schr\"{o}dinger equation with the Hamiltonian~\eqref{eqn:Hamiltoian_dimless_ex} is presented in Fig.~\ref{fig:app_ex_wf}. The corresponding wave function of the original Hamiltonian is simply given by $\varphi_{x}(\rho)=(\eta/a_{\textup{B}})\phi(\rho/a_{\textup{B}})$, where the numerical value of $\eta$ equals $0.615$.

\subsection{Biexciton}

The Hamiltonian of the two holes and two electrons moving within the TMD quantum well has the form
\begin{multline}\label{eqn:Hamiltoian_dim}
   H_b=-\dfrac{\hbar^2}{2m_e}\left(\Delta_e+\Delta_{e'}\right)-\dfrac{\hbar^2}{2m_h}\left(\Delta_h+\Delta_{h'}\right)\\
   -V_{eh}-V_{eh'}-V_{e'h}-V_{e'h'}+V_{ee'}+V_{hh'}.
\end{multline}
Using the dimensionless units introduced earlier, we can rewrite the Hamiltonian in a dimensionless form as follows:
\begin{multline}
    \tilde{H}_b=-\dfrac{\sigma}{1+\sigma}\left(\tilde{\Delta}_e+\tilde{\Delta}_{e'}\right)-\dfrac{1}{1+\sigma}\left(\tilde{\Delta}_h+\tilde{\Delta}_{h'}\right)\\
    -2\left(\tilde{V}_{eh}+\tilde{V}_{eh'}+\tilde{V}_{e'h}+\tilde{V}_{e'h'}-\tilde{V}_{ee'}-\tilde{V}_{hh'}\right),
\end{multline}
where $\sigma=m_e/m_h$.

Utilizing the Hamiltonian~\eqref{eqn:Hamiltoian_dim} and biexciton wave function~\eqref{eq:bie_wf}, one can calculate the binding energy of the four-particle system via
\begin{align}\label{eqn:total_binding_energy_of_4_particles}
    B_{2e,2h}=\!-\!\!\int{\rm d}\Omega \psi_b({\bf r}_e,\! {\bf r}_e',\!{\bf r}_h,\!{\bf r}_h')^*H_b \psi_b({\bf r}_e,\! {\bf r}_e',\!{\bf r}_h,\!{\bf r}_h'),\!\!
\end{align}
where the center-of-mass momentum ${\bf K}_b$ is equal to zero and the total volume of integration is defined as  ${\rm d}\Omega={\rm  d}{\bf r}_e{\rm  d}{\bf r}_{e'}{\rm  d}{\bf r}_h{\rm  d}{\bf r}_{h'}$. By maximizing the value of $B_{2e,2h}$, we can determine the correct form of the biexciton wave function, which we use to calculate the constant $V_b$. Taking into account~\eqref{eq:bie_wf}, the only remaining issue concerns how to determine the function $\psi({\bf r},{\bf r}',{\bf R})$ describing the relative motion of the particles constituting the biexciton.  As stated in the main text, we use the variational ansatz proposed in~\cite{Kleinman1983}, which is applicable specifically for biexciton particles. The corresponding function up to normalization can be found in the following form:
\begin{multline}
    \varphi_b\!(k,s_1,s_2,t_1,t_2,v)\!=\!\left[ (kv)^{\nu}\!\exp{\left(-\rho k v\right)}+\lambda\exp{\left(-\tau k v \right)}\right]\\ \times\exp{\left[-\dfrac{k(s_1+s_2)}{2}\right]}\cosh{\left[\dfrac{\beta k (t_1-t_2)}{2}\right]},
\end{multline}
where variables $s_1$, $s_2$, $t_1$, $t_2$,  and $v$ were defined in the main text. It is easy to show that the normalized version of such a wave function $ \varphi_b^N$ is related to the original one $\psi({\bf r},{\bf r}',{\bf R})$ as follows:
\begin{widetext}
\begin{multline}\label{eqn:connection_formula}
    \psi({\bf r},{\bf r}',{\bf R})\!=\!\varphi_b^{N}\!\!\left(k,\!\left|{\bf R}\!-\!\dfrac{{\bf  r}\!-\!{\bf  r}'}{2}\!\right|\!+\!\left|{\bf R}\!-\!\dfrac{{\bf  r}\!+\!{\bf r}'}{2}\right|\!,\!\left|{\bf R}\!+\!\dfrac{{\bf  r}\!+\!{\bf  r}'}{2}\right|\!+\!\left|{\bf R}\!+\!\dfrac{{\bf  r}\!-\!{\bf r}'}{2}\right|\!\right.,\\ \left.\left|{\bf R}\!-\!\dfrac{{\bf  r}\!-\!{\bf  r}'}{2}\right|\!-\!\left|{\bf R}\!-\!\dfrac{{\bf  r}\!+\!{\bf r}'}{2}\right|\!,\!\left|{\bf R}\!+\!\dfrac{{\bf  r}\!+\!{\bf  r}'}{2}\right|\!-\!\left|{\bf R}\!+\!\dfrac{{\bf  r}\!-\!{\bf r}'}{2}\right|\!,\!|{\bf r}'|\!\right).
\end{multline}
\end{widetext}
Let us first compute the normalization constant for the biexciton wave function:
\begin{multline}
    N_b=\int{\rm  d}{\bf r}_e{\rm  d}{\bf r}_{e'}{\rm  d}{\bf r}_h{\rm  d}{\bf r}_{h'}\dfrac{\exp{\left[i{\bf R}_b({\bf K}_b-{\bf K}'_b)\right]}}{S}\\
    \times\left.\varphi_b(k,s_1,s_2,t_1,t_2,v)^2\right\rvert_{{\bf K}_b={\bf K}'_b}.
\end{multline}
By changing variables from ${\bf r}_e$, ${\bf r}_{e'}$, ${\bf r}_h$, ${\bf r}_{h'}$ to ${\bf r}_{hh'}$, ${\bf r}_{eh}$, ${\bf r}_{e'h}$, ${\bf R}_b$ and taking into account the trivial dependence of the wave function on ${\bf R}_b$, we obtain
\begin{multline}
    N_b=2\pi\int r_{hh'}{\rm  d}r_{hh'}r_{eh}{\rm  d}r_{eh}r_{e'h}{\rm  d}r_{e'h}{\rm  d}\phi_{eh}{\rm  d}\phi_{e'h} \\ \times \varphi_b(k,s_1,s_2,t_1,t_2,v)^2,
\end{multline}
where we rewrite the integrand within polar coordinates and factor $2\pi$ appears due to the fact that the system is invariant with respect to a common rotation, which allows one to reduce the number of integrations from six to five. Moving finally to the variables $s_1$, $s_2$, $t_1$, $t_2$,  and $v$, we find
\begin{multline}
    N_b=8\pi\int\limits_0^{\infty} v{\rm  d}v \int\limits_v^{\infty}{\rm  d}s_1\int\limits_{-v}^{v}{\rm  d}t_1 \int\limits_v^{\infty}{\rm  d}s_2\int\limits_{-v}^{v}{\rm  d}t_2 
    \\ \times J_{st}(s_1,t_1,v) J_{st}(s_2,t_2,v)
   \varphi_b(k,s_1,s_2,t_1,t_2,v)^2,
\end{multline}
where Jacobians $J_{st}$ associated with changes of variables are defined by
\begin{align}
    J_{st}(s,t,v)=\dfrac{1}{4}\dfrac{s^2-t^2}{\sqrt{s^2-v^2}\sqrt{v^2-t^2}}.
\end{align}
It turns out that four integrations can be performed analytically. Changing then the variables according to $\tilde{k} x_v \rightarrow x_v$, we obtain the following final expression:
\begin{align}
    N_b=\dfrac{\pi^3}{4 k^6}\int\limits_0^{\infty} {\rm  d}x_v \  x_v^3 \chi^2(T_2^2+T_4^2),
\end{align}
where, following~\cite{Kleinman1983}, we have introduced
\begin{align}
    \chi&=x_v^{\nu}\exp{\left(-\rho x_v\right)}+\lambda \exp{\left(-\tau x_v\right)},\\
    T_2&=K_1(x_v)+v K_0(x_v)/2,\\
    T_4&=K_1(v)I_0(\beta x_v)+v K_0(v)I_1(\beta x_v)/\beta.
\end{align}
It is also convenient to introduce a dimensionless quantity defined as:
\begin{align}\label{eqn:normalization_tilde}
    \tilde{N}_b=k^6 N_b.
\end{align}
Now let us calculate the total binding energy $B_{2e,2h}$ of the four-particle system taking into account the choice of the variational function. Partly using the notation of the work~\cite{Kleinman1983}, we introduce a dimensionless counterpart of $B_{2e,2h}$:
\begin{align}\label{eqn:total_be_biex_dimless}
    \tilde{B}_{2e,2h}=\dfrac{2\tilde{L}^{\textup{KR}}(\tilde{k})-\tilde{k}^2\tilde{M}}{\tilde{N}},
\end{align}
where we explicitly indicate which terms contain the dependence on the variational parameter $\tilde{k}$. The
functions $\tilde{N}$, $\tilde{M}$, and $\tilde{L}^{\textup{KR}}$ are defined as follows:
\allowdisplaybreaks
\begin{align}
    &\tilde{N}\!=\!\int\limits_0^{\infty}\! {\rm  d}x_v x_v \chi^2\left[x_v^2(T_2^2+T_4^2)\right],\\
    &\tilde{M}\!=\!\int\limits_0^{\infty}\!{\rm  d}x_v x_v \chi^2\big\{x_v^2(T_1 T_2+2 T_3 T_4)+x_v^2\left[\sigma/(1+\sigma)\right]\nonumber\\
    &\ \ \ \times \left[(T_2^2+T_4^2)(\chi'/\chi)^2-4(T_2T_8+T_4T_7)(\chi'/\chi)\right.\\
    &\qquad\left.+(1+\beta^2)(T_8^2-T_2T_6)+T_7^2-(1-\beta^2)T_4T_5\right]\big\},\nonumber\\
    &\tilde{L}^{\textup{KR}}(\tilde{k})\!=\!\!\int\limits_0^{\infty}\!\! {\rm  d}x_v x_v \chi^2\!\left[q_{4eh}(\tilde{k})\!-\!q_{hh'}(\tilde{k})\!-\!q_{ee'}(\tilde{k})\right],\!\!\\
    &q_{4eh}(\tilde{k})=\dfrac{4 x_v}{\pi\beta}\!\int\limits_{x_v}^{\infty}\!\!{\rm d}x_s\!\!\!\int\limits_{-x_v}^{x_v}\!\!{\rm d}x_t J_{st}(x_s,x_t,x_v)\nonumber\\
    &\ \ \times\exp{\left(-x_s\right)}\left[\tilde{V}\left(\dfrac{x_s+x_t}{2\tilde{k}}\right)+\tilde{V}\left(\dfrac{x_s+x_t}{2\tilde{k}}\right)\right]\nonumber\\
    &\ \ \times \{2\cosh{\left(-\beta x_t\right)}\left[I_1(\beta x_v)K_0(x_v)\right.
    \nonumber\\
    &\qquad\qquad\quad+\left.\beta I_0(\beta x_v)K_1(x_v)\right]+\beta x_v K_2(x_v)\},\\
    &q_{hh'}(\tilde{k})=x_v^2(T_2(x_v)^2+T_4(x_v)^2)\tilde{V}\left(x_v/\tilde{k}\right),\\
    &T_1=2K_1(x_v)-\beta^2 x_v K_0(x_v),\\
    &T_3=K_1(x_v)I_0(\beta x_v)+\beta K_0(x_v)I_1(\beta x_v),\\
    &T_5=K_1(x_v)I_0(\beta x_v)- K_0(x_v)I_1(\beta x_v)/\beta,\\
    &T_6=K_1(x_v)- x_v K_0(x_v)/2,\\
    &T_7=(1/\beta-\beta)K_1(x_v)I_1(\beta x_v),\\
    &T_8=x_v K_1(x_v)/2.
\end{align}
The function $q_{ee'}(\tilde{k})$ associated with the Hamiltonian term $\tilde{V}_{ee'}\equiv \tilde{V}(x_u/\tilde{k})$ requires a separate treatment. In contrast to the strategy chosen in~\cite{Kleinman1983}, we do not address extrapolations used there and calculate this function exactly within the accuracy of numerical integration. To calculate the term $q_{ee'}(\tilde{k})$, we have to use a different set of variables. First, by a linear transformation $s_1 = r_{eh} + r_{eh'}$, $t_1 = r_{eh} - r_{eh'}$, $s_2 = r_{e'h} + r_{e'h'}$, $t_2 = r_{e'h} - r_{e'h'}$, $u = r_{ee'}$, $v = r_{hh'}$, we express the function $\varphi_b$ in terms of $r_{eh},r_{eh'},r_{e'h},r_{e'h'}$ instead of $s_1$, $s_2$, $t_1$, $t_2$. Next, we exclude $r_{eh'}$ and $r_{e'h'}$ according to $v=r_{hh'}$, $u=r_{ee'}$, $r_{eh}$, $r_{e'h}$ and an additional angle $\theta'$:
\begin{align}\label{eqn:}
&r_{e'h'}=\sqrt{v^2-2vr_{e'h}\cos{\theta'}+r_{e'h}^2},\\
&r_{e'h'}=\sqrt{v^2-2vr_{eh}\cos{\theta''}+r_{eh}^2},\\
&\theta''=\theta'+\arccos{\left(\dfrac{r_{eh}^2+r_{e'h}^2-u^2}{2r_{eh}r_{e'h}}\right)}.
\end{align}
Finally, we replace $r_{e'h}$ and $r_{eh}$ by $s_a=r_{eh}+r_{e'h}$ and $t_a=r_{eh}-r_{e'h}$. Thus, the expression for $q_{ee'}(\tilde{k})$ can be recast into
\begin{align}
&q_{ee'}(\tilde{k})=\dfrac{4}{\pi^2}\int\limits_0^{\infty}{\rm d}x_u \left[g_{ee'} x_u \tilde{V}\left(x_u/\tilde{k}\right) \right],
\end{align}
where the function $g_{ee'}(x_u)$ does not depend on $\tilde{k}$ and is defined as
\begin{multline}
g_{ee'}=\int\limits_{0}^{2\pi}{\rm d}\theta'\int\limits_{x_u}^{\infty}{\rm d} x_{s_a}\!\!\!\!\int\limits_{-x_u}^{x_u}\!\! {\rm d} x_{t_a}J_{st}(x_{s_a},x_{t_a},x_u)\\
\times \Upsilon(x_{s_1},x_{s_2},x_{t_1},x_{t_1})^2.
\end{multline}
Here, the part $\Upsilon$ of the relative-motion biexciton wave function $\varphi_b$ reads
\begin{multline}
    \Upsilon(x_{s_1},x_{s_2},x_{t_1},x_{t_1})=\exp{\left[-\dfrac{x_{s_1}+x_{s_2}}{2}\right]}\\\times 
    \cosh{\left[\dfrac{\beta\left(x_{t_1}-x_{t_2}\right)}{2}\right]}.
\end{multline}
In this function, of course, before integration, the change of variables described above should be done. From the computational point of view, $g_{ee'}$ is a function of the dimensionless coordinates $x_v$, $x_u$, and variational parameter $\beta$. To optimize the process of finding the maximum value of the total binding energy $\tilde{B}_{2e,2h}$ by varying the parameters $\lambda$, $\tau$, $\rho$, $\nu$, we construct interpolation polynomials for $g_{ee'}(x_u)$ and then interpolation polynomials for $q_{ee'}(\tilde{k})$, which is a function of $x_v$, $\beta$, and $\tilde{k}$. Note also that $g_{ee'}$ and $q_{ee'}$ are also functions of $x_0$, but since it is a constant for a specific material (WS$_2$), we do not focus on this dependence. 

We described all of the steps that are necessary for computing the dimensionless total binding energy~\eqref{eqn:total_be_biex_dimless}. By varying the values of the parameters $\tilde{k}$, $\beta$, $\nu$, $\rho$, $\lambda$, and $\tau$, we found the following maximum value of the total binding energy: $\tilde{B}_{2e,2h}=2.985$, which then yields $B_{2e,2h}=313~\textup{meV}$.

\section{Dimensionless system of equations}\label{appendix:dimensionless}
Here we present a dimensionless analogue of the system \addMisha{\eqref{motion}} that we used in the practical calculations. By introducing dimensionless quantities $e_c=1$, $e_x=E_x/E_c$, $e_b=2$, $v_x = V_x /E_c$, and measuring time  in units of $\hbar/E_c$, we obtain the following system:
\begin{subequations}
\label{motion:appendix}
\begin{align}
\mathrm i \frac{{\rm d}\tilde{C}_\pm}{{\rm d}\tilde{t}} =& e_c \tilde{C}_\pm + v_x \tilde{X}_\pm + \alpha v_x \tilde{X}_\mp^* \tilde{B} ,\label{si_eqn:sys_eq_12}\\
\mathrm i \frac{{\rm d} \tilde{X}_\pm}{{\rm d}\tilde{t}} =& e_x \tilde{X}_\pm + v_x \tilde{C}_\pm + \alpha v_x  \tilde{C}_\mp^* \tilde{B},\label{si_eqn:sys_eq_34}\\
\mathrm i \frac{{\rm d} \tilde{B}}{{\rm d}\tilde{t}} =& e_b \tilde{B} + \alpha  v_x (\tilde{X}_-\tilde{C}_+ + \tilde{X}_+ \tilde{C}_-),\label{si_eqn:sys_eq_5}
\end{align}
\end{subequations}
where $\tilde{C}$ is introduced in such a way that $|\tilde{C}|^2$ gives the number of particles in an area of $a_{\textup{B}}^2$. In the case of WS$_2$, we obtain $e_x=1.0154$, $v_x = 0.0111$. Moreover, by shifting the energy for photons and excitons by $E_c$ [$\tilde{C}_\pm\rightarrow \tilde{C}_\pm \exp{(-i e_c\tilde{t})}$  and  $\tilde{X}_\pm\rightarrow \tilde{X}_\pm \exp{(-i e_c\tilde{t})}$], and the energy for the biexciton by $2E_c$ [$\tilde{B}\rightarrow \tilde{B} \exp{(-i 2e_c\tilde{t})}$], the system can be rewritten so that it depends only on two parameters, $\alpha$ and $\kappa=(e_x-e_c)/v_x$.


\bibliography{biexciton}
\end{document}